\documentclass[twocolumn,showpacs,preprintnumbers,amsmath,amssymb]{revtex4}

\usepackage{amsmath,amssymb,graphicx}
\usepackage{graphicx}
\usepackage{dcolumn}
\usepackage{bm}
\usepackage{epsfig}
\usepackage{here}
\usepackage{float}
\usepackage{amsmath}
\usepackage[usenames]{color}

\newcommand{\bea}{\begin{eqnarray}}
\newcommand{\eea}{\end{eqnarray}}

\begin{document}
\draft

\title{On cosmologically designed modified gravity theories}
\author{Jai-chan Hwang${}^{1}$, Hyerim Noh${}^{2}$, and Chan-Gyung Park${}^{1}$, }
\address{${}^{1}$Department of Astronomy and Atmospheric Sciences,
                 Kyungpook National University, Taegu, Korea \\
         ${}^{2}$Korea Astronomy and Space Science Institute,
                 Daejon, Korea}


\begin{abstract}

Versions of parameterized pseudo-Newtonian gravity theories
specially designed for cosmology have been introduced in recent
cosmology literature. The modifications demand a zero-pressure fluid
in the context of versions of modified Poisson-like equation with
two different gravitational potentials. We consider such
modifications in the context of relativistic gravity theories where
the action is a general algebraic function of the scalar curvature,
the scalar field, and the kinetic term of the field. In general it
is not possible to isolate the zero-pressure fluid component
simultaneously demanding a modification in the Poisson-like
equation. Only in the small-scale limit we can realize some special
forms of the attempted modifications. We address some loopholes in
the possibility of showing non-Einstein gravity nature based on
pseudo-Newtonian modifications in the cosmological context. We point
out that future observations of gravitational weak lensing together
with velocity perturbation can potentially test the validity of
Einstein's gravity in cosmology context.

\end{abstract}

\noindent 

\maketitle

\section{Introduction}

In parallel with recent growth of cosmologically relevant
observational data, theoretical cosmology also has been growing
rapidly. Theories meet with observations mostly in the linearly
perturbed stage of the large-scale cosmological structures. In our
present understanding, the simple theoretical model of currently
$\Lambda$CDM (cosmological constant $\Lambda$ as the dark energy
together with the cold dark matter) dominated model, together with
inflation generated initial fluctuations by a single field, has
enough parameter space to amply cope with current observations.
Recently, however, motivated by potential future precision data,
more complicated models for the dark energy sector have been pursued
in the literature. These include a fluid model with exotic equation
of states, a scalar field with diverse potentials, modified
relativistic gravity theories, etc; see Ref.\ \cite{DE-review-2010}
for recent reviews.

One recent attempt concerns modifying perturbation equations by some
unknown parameters in the level of modified gravity or
pseudo-Newtonian gravity \cite{Acquaviva-etal-2008,
Amendola-etal-2008,Bean-2009,Bertschinger-Zukin-2008,Beynon-etal-2009,
Caldwell-etal-2007,Cui-etal-2010,Daniel-etal-2008,Daniel-etal-2009,Daniel_etal-2010,
DeFelice-Suyama-2010,
DeFelice-etal-2010,Dore-etal-2007,Ferreira-2010,Giannantonio-etal-2009,
Guzik-etal-2009,Hu-2008,Hu-Sawicki-2007,Huterer-Linder-2007,Jain-Zhang-2008,
Laszlo-Bean-2008,Linder-Cahn-2007,Martinelli-etal-2010,Nesseris-2007,Pogosian_etal-2010,
Schmidt-2008,Serra-etal-2009,Shapiro-etal-2010,Skordis-2009,Song-Koyama-2008,
Song-Dore-2009,Song-etal-2010-01,Song_etal-2010-11,Tsujikawa-2007,Tsujikawa-etal-2008,
Uzan-2010,
Zhang-etal-2008,Zhang-etal-2007,Zhao-etal-2010,Zhao-etal-2009}.
Effects of the parameters on the cosmic microwave background
radiation,
the weak lensing shear field
and the growth of large-scale structures,
have been studied, and constraints on the parametrized modified
gravity models based on current and future observational data have
been given. Proper theoretical justification of such attempts would
be desirable. One way of justifying the situation is to show that
such modifications are allowed or can be implemented in the
relativistic gravity theories based on the action formulation. In
this work we will examine $f(R, \phi, X)$ gravity as a potential
candidate. The $f(R, \phi, X)$ gravity is presented as an action
\bea
   & & S = \int d^4 x \sqrt{-g} \left[
       {1 \over 2} f (R, \phi, X) + L_m \right],
\eea where $f$ is a general algebraic function of the scalar
curvature $R$, a scalar field $\phi$ and the kinetic combination $X
\equiv {1 \over 2} \phi^{,c} \phi_{,c}$; $L_m$ is the matter
Lagrangian. This includes $f(R)$ gravity and many other gravity
theories like Brans-Dicke theory, scalar-tensor theory,
non-minimally coupled scalar field theories, etc, as cases.

We will show that in general the pseudo-Newtonian modifications
cannot be accommodated in our considered generalized gravity
theories, see Sec.\ \ref{sec:validity}. Only in the small scales far
inside the horizon we often have a closed form second-order
differential equation of CDM (or baryon) density perturbation in a
certain gauge condition with specific effective gravitational
constant, see Sec.\ \ref{sec:small-scale}. Such small scale
coincidences in several modified gravity theories may have motivated
the pseudo-Newtonian approach. We point out some loopholes in the
pseudo-Newtonian approach, see Sec.\ \ref{sec:loopholes}. Even in
Einstein's gravity theory with a minimally coupled scalar field as a
dynamical dark energy, we can easily introduce the field potential
where the dynamical nature of dark energy is important in
perturbation evolution. We believe it is desirable to explain the
observational data based on theoretically motivated models, and also
based on proper treatment of the complete set of equations provided
in such models. In Sec.\ \ref{sec:loopholes} we also point out
future observations which can be used to test potentially Einstein
gravity nature.

The Appendix presents the complete set of equations in the $f(R, \phi,
X)$ gravity in the presence of additional matter. These will be
useful in future numerical study of perturbation evolution in such
gravity theories. We set $c \equiv 1 \equiv 8 \pi G$.

\section{Validity of cosmologically modified pseudo-Newtonian gravity}
                                           \label{sec:validity}

Several versions of phenomenologically modified gravity theories
specially designed for cosmological purpose have been introduced in
the literature
\cite{Acquaviva-etal-2008,Amendola-etal-2008,Bertschinger-Zukin-2008,
Bean-2009,Caldwell-etal-2007,Cui-etal-2010,Daniel-etal-2008,Daniel-etal-2009,
Daniel_etal-2010,Dore-etal-2007,Ferreira-2010,Giannantonio-etal-2009,
Guzik-etal-2009,Hu-2008,Hu-Sawicki-2007,Jain-Zhang-2008,Laszlo-Bean-2008,
Martinelli-etal-2010,Pogosian_etal-2010,Schmidt-2008,Serra-etal-2009,
Shapiro-etal-2010,Song-Koyama-2008,
Song-etal-2010-01,Song_etal-2010-11,Zhang-etal-2007,Zhang-etal-2008,
Zhang-etal-2008,Zhao-etal-2009}. The modified gravity {\it assumes}
\bea
   & & \varphi_\chi = - \bar \eta \alpha_\chi,
   \label{mod-1} \\
   & & {k^2 \over a^2} \alpha_\chi
       = - {1 \over 2} \bar \mu \delta \mu_v,
   \label{mod-2}
\eea where $\varphi_\chi \equiv \varphi - H \chi$ and $\alpha_\chi
\equiv \alpha - \dot \chi$ are gauge-invariant combinations which
are the same as $\varphi$ and $\alpha$, respectively, in the
zero-shear gauge ($\chi \equiv 0$); $\delta \mu_v \equiv \delta \mu
- \dot \mu a v$ is a gauge-invariant combination which is the same
as $\delta \mu$ in the comoving gauge ($v \equiv 0$). In Bardeen's
notation \cite{Bardeen-1980} we have $\varphi_\chi = \Phi_H$,
$\alpha_\chi = \Phi_A$, and $\delta \mu_v = \mu \epsilon_m$. The
gauge-invariant combinations $\alpha_\chi$ and $\varphi_\chi$ can be
interpreted as the Newtonian and the post-Newtonian potentials,
respectively \cite{PN}. The variables $\bar \eta = \bar \eta (k, t)$
and $\bar \mu = \bar \mu (k, t)$ are new free parameters introduced
to modify Einstein's gravity where $\bar \eta \equiv 1 \equiv \bar
\mu$. Later we will show that we could have $\bar \mu \neq 1$ even
in Einstein's gravity.

Our metric and energy-momentum tensor conventions are \bea
   & & ds^2 = - \left( 1 + 2 \alpha \right) a^2 d \eta^2
       - 2 a^2 \beta_{,\alpha} d \eta d x^\alpha
   \nonumber \\
   & & \quad
       + a^2 \left[ g^{(3)}_{\alpha\beta} \left( 1 + 2 \varphi \right)
       + 2 \gamma_{,\alpha|\beta} \right]
       dx^\alpha dx^\beta,
   \label{metric}
\eea with $\chi \equiv a \beta + a^2 \dot \gamma$, and \bea
   & & T^0_0 = - \mu - \delta \mu, \quad
       T^0_\alpha = - ( \mu + p ) {1 \over k} v_{,\alpha},
   \nonumber \\
   & &
       T^\alpha_\beta
       = ( p + \delta p ) \delta^\alpha_\beta
       + {1 \over a^2} \left( \Pi^{|\alpha}_{\;\;\; \beta}
       - {1 \over 3} \delta^\alpha_\beta \Pi^{|\gamma}_{\;\;\; \gamma} \right).
   \label{Tab}
\eea A vertical bar is a covariant derivative based on the comoving
three-space metric $g^{(3)}_{\alpha\beta}$ of the Robertson-Walker
spacetime. The complete sets of scalar-type perturbation equations
in Einstein's gravity and in the $f(R, \phi, X)$ gravity are
presented in the Appendix A and B, respectively.

It is important to notice that, in the presence of multiple
component of fluids or a fluid in the context of generalized
relativistic gravity theories, the fluid quantities appearing in
Eqs.\ (\ref{mod-2}) and (\ref{Tab}) are collective ones, see Eqs.\
(\ref{fluid-sum}) in Einstein's gravity, and Eqs.\
(\ref{effective-fluid}) and (\ref{fluid-sum-X}) in the $f(R, \phi,
X)$ gravity. In the following, except in the Appendix A and B, we
{\it assume} a flat background, thus set $K = 0$.

\subsection{Single component situation}

We consider a single component in an yet unspecified generalized
gravity, and accept \cite{Pogosian_etal-2010,Daniel_etal-2010} \bea
   & & \dot \delta_{\chi}
       = - {k \over a} v_{\chi}
       - 3 \dot \varphi_\chi,
   \\
   & & \dot v_{\chi} + H v_{\chi}
       = {k \over a} \alpha_\chi,
\eea where $\delta \equiv \delta \mu/\mu$. These equations follow
from Eqs.\ (\ref{eq-6}) and (\ref{eq-1}), and Eq.\ (\ref{eq-7}),
respectively, by {\it demanding} $p = 0$ and $\delta p = 0 = \Pi$.
As we consider a single component, all fluid quantities are about
that single fluid supported by the nature of generalized gravity. In
this case, because we assume $\Pi = 0$, from Eq.\ (\ref{eq-4}), and
Eqs.\ (\ref{eq-2}) and (\ref{eq-3}), respectively, we inevitably end
up with \bea
   & & \varphi_\chi = - \alpha_\chi,
   \label{varphi-alpha-Pogosyan} \\
   & & {k^2 \over a^2} \alpha_\chi
       = - {1 \over 2} \delta \mu_v.
\eea In the $f(R, \phi, X)$ gravity context we have $\mu = \mu_X$,
$p = p_X$, $\delta \mu = \delta \mu_X$, $\delta p = \delta p_X$.
$(\mu + p) v = (\mu_X + p_X) v_X$, and $\Pi = \Pi_X$; the subindex
$X$ indicates the contribution from $f(R, \phi, X)$ gravity sector;
the fluid quantities of the $X$-component are presented in Eqs.\
(\ref{fluid-BG}) and (\ref{fluid-X}). This apparently implies $\bar
\eta = 1 = \bar \mu$. That is, in the context of our generalized
relativistic gravity theories the phenomenological modifications in
Eqs.\ (\ref{mod-1}) and (\ref{mod-2}) with $\bar \eta \neq 1$ or
$\bar \mu \neq 1$ are {\it not} allowed.

Even in the case such a fluid with $\bar \mu = 1 = \bar \eta$ is
implemented in the context of $f(R, \phi, X)$ gravity, thus $\mu =
\mu_X \propto a^{-3}$ etc, those are at best uninteresting, because
the nontrivial solutions in $f(R, \phi, X)$ gravity should not
affect the Einstein's gravity nature of the metric and fluid
variables.

\subsection{Multi-component situation}

Now we consider a CDM component in an yet unspecified generalized
gravity theory, and accept \bea
   & & \dot \delta_{c \chi}
       = - {k \over a} v_{c \chi}
       - 3 \dot \varphi_\chi,
   \label{Bean-eq1} \\
   & & \dot v_{c \chi} + H v_{c \chi}
       = {k \over a} \alpha_\chi,
   \label{Bean-eq2}
\eea where a subindex $c$ indicates the CDM component. Equations (\ref{Bean-eq1}) and
(\ref{Bean-eq2}) follow from Eqs.\ (\ref{eq-6i}) and (\ref{eq-1}),
and Eq.\ (\ref{eq-7i}), respectively, by {\it demanding} $p_c = 0$
and $\delta p_c = 0 = \Pi_c$. In this case, from Eq.\ (\ref{eq-4}),
and Eqs.\ (\ref{eq-2}) and (\ref{eq-3}), respectively, we have \bea
   & & \varphi_\chi = - \alpha_\chi - \Pi,
   \label{varphi-alpha-Bean} \\
   & & {k^2 \over a^2} \varphi_\chi
       = {1 \over 2} \delta \mu_v.
   \label{Poisson-eq}
\eea Thus, Eqs.\ (\ref{mod-1}) and (\ref{mod-2}) {\it demand} $\Pi =
( \bar \eta - 1 ) \alpha_\chi$ and \bea
   & & \bar \mu = {1 \over \bar \eta}.
   \label{Pi-alpha_chi}
\eea This is apparently a stringent relation required between the
two factors introduced in Eqs.\ (\ref{mod-1}) and (\ref{mod-2}).

If a CDM component is present in the context of $f(R, \phi, X)$
gravity, from Eqs.\ (\ref{effective-fluid}) and (\ref{fluid-X}) we
have $\Pi = \Pi_X = \delta F_\chi/F$; $\delta F_\chi \equiv \delta F
- \dot F \chi$ is a gauge-invariant combination which is the same as
$\delta F$ in the zero-shear guage ($\chi \equiv 0$); we have $F
\equiv
\partial f/\partial R$. Thus, Eq.\ (\ref{Pi-alpha_chi}) {\it
demands} \bea
   & & {1 \over F} \delta F_\chi = ( \bar \eta - 1 ) \alpha_\chi.
\eea This is a strong physical condition on the nature of a
gauge-invariant combination $\delta F_\chi$.

More seriously, however, it is important to notice that $\delta
\mu_v$ in Eq.\ (\ref{Poisson-eq}) is a collective density
perturbation in the collective comoving gauge ($v \equiv 0$); i.e.,
we have \bea
   & & \delta \mu_v \equiv \delta \mu + 3 {a \over k} H (\mu + p) v,
   \label{delta-mu_v-def}
\eea where $\delta \mu$ and $(\mu + p) v$ are sum of the fluid
quantities, see Eqs.\ (\ref{fluid-sum}) and (\ref{fluid-sum-X}).
On the other hand, the CDM density perturbation in the CDM-comoving gauge is
$\delta \mu_{c v_c} \equiv \delta \mu_c + 3 (a/k) H \mu_c v_c = \delta \mu_{c \chi} +
3 (a/k) H \mu_c v_{c \chi}$. If one of the components is due to
generalized gravity theory like $f(R, \phi, X)$ gravity, the
combination becomes more complicated. If we consider the CDM in
$f(R, \phi, X)$ gravity, from Eq.\ (\ref{effective-fluid}) we have
\bea
   & & \delta \mu = {1 \over F} \delta \mu_c
       + \delta \mu_X - {\mu_c \over F^2} \delta F,
   \nonumber \\
   & & \left( \mu + p \right) v
       = {1 \over F} \mu_c v_c
       + \left( \mu_X + p_X \right) v_X,
   \nonumber \\
   & &
       \delta p = \delta p_X, \quad
       \Pi = \Pi_X.
   \label{effective-fluid-Bean}
\eea Thus, we have \bea
   & & \delta \mu_v
       = {1 \over F} \delta \mu_{c v_c}
       + \delta \mu_{X v_X}
       - {\mu_c \over F^2} \delta F_{v_X},
   \label{delta-mu_v}
\eea where \bea
   & & \delta \mu_{c v_c}
       \equiv \delta \mu_c - \dot \mu_c {a \over k} v_c, \quad
       \delta \mu_{X v_X}
       \equiv \delta \mu_X - \dot \mu_X {a \over k} v_X,
   \nonumber \\
   & & \delta F_{v_X}
       \equiv \delta F - \dot F {a \over k} v_X.
\eea Therefore, unless we have $F = 1$ and $\delta \mu_{X v_X} -
(\mu_c / F^2) \delta F_{v_X} =0$, both of which are quite nontrivial
physical conditions on the nature of the generalized gravity, it is
not possible to interpret $\delta \mu_v$ in Eq.\ (\ref{mod-2}) as
$\delta \mu_{c v_c}$ which is $\delta \mu_c$ in the CDM-comoving
gauge ($v_c \equiv 0$). It is not likely that after implementing
these conditions we have meaningful nontrivial results. For example,
we can show that, in the $f(R)$ gravity case, implementing $F = 1$,
$\delta \mu_X - (\mu_c/F^2) \delta F = 0$ and $v_X = 0$ simply leads
to an inconsistent equation for the background, thus impossible to
implement.

Later, in the small-scale limit, in  Eq.\ (\ref{delta-mu_v-small})
we will show \bea
   & & \delta \mu_v
       \simeq {1 \over F} \delta \mu_{c v_c}
       + {\bar \eta - 1 \over \bar \eta + 1}
       {1 \over F} \delta \mu_{c v_c}.
\eea Thus, compared with Eq.\ (\ref{delta-mu_v}), apparently the
contribution from $X$-component is important in our generalized
gravity theory.

In the context of generalized gravity theories, in general, it is
unavoidable (except for the cosmological constant case) that we have
additionally coupled second-order differential equation for the
$X$-component representing the $f(R, \phi, X)$ gravity sector. The
complete sets of equations of CDM perturbation in the context of
$f(R, \phi, X)$ gravity, which inevitably lead to a fourth-order
differential equation, are presented in the Appendix C. In the
zero-shear gauge Eqs.\ (\ref{Bean-eq1}) and (\ref{Bean-eq2})
together with Eqs.\ (\ref{Bean-eq3}) and (\ref{Bean-eq4}) are the
complete sets we need. In the Appendix C we also present the
complete sets of equations in the uniform-field gauge and the
CDM-comoving gauge. In all cases we end up having fourth-order
differential equations.

\section{Small-scale limit approximation}
                                    \label{sec:small-scale}

In the literature we often find a second-order differential equation
of $\delta_{c v_c}$ or $\delta_{c \chi}$ derived in the small-scale
limit \cite{Tsujikawa-2007,Tsujikawa-etal-2008,DeFelice-etal-2010}.
By naively preserving only terms attached with $k^2$ except for
$\delta_{c}$ term, Eq.\ (\ref{ddot-delta-eq}) becomes \bea
   \ddot \delta_{c v_c} + 2 H \dot \delta_{c v_c}
       \simeq - {k^2 \over a^2} \alpha_\chi.
   \label{ddot-delta-eq-CCG-SS}
\eea From Eqs.\ (\ref{eq-4}) and (\ref{fluid-X}) we have \bea
   \varphi_\chi + \alpha_\chi = - {\delta F_\chi \over F}.
   \label{varphi_chi-SS1}
\eea In the small-scale limit from Eq.\ (\ref{Poisson-eq}), using
Eqs.\ (\ref{effective-fluid-Bean}), (\ref{fluid-X}) and
(\ref{eq-1}), we have \bea
   {k^2 \over a^2} \left( \varphi_\chi
       + {\delta F_\chi \over 2 F} \right)
       \simeq {1 \over 2 F} \delta \mu_{c v_c}.
   \label{varphi_chi-SS2}
\eea We need one more relation. From the equation of motion in Eq.\
(\ref{EOM-pert-f}) and using Eqs.\ (\ref{delta-R}),
(\ref{varphi_chi-SS1}), and $\delta F = F_{,\phi} \delta \phi +
F_{,R} \delta R$ [here we {\it consider} $F_{,X} = 0$ case] we can derive
\bea
   \delta R_\chi
   &\simeq& - {1 \over F_{,\phi}} \left( f_{,\phi\phi}
       + f_{,X} {k^2 \over a^2} \right) \delta \phi_\chi
   \nonumber \\
   &\simeq& {- 2 {k^2 \over a^2} \left( \alpha_\chi
       + 2 {F_{,\phi} \over F} \delta \phi_\chi \right)
       \over
       1 + 4 {F_{,R} \over F} {k^2 \over a^2} },
   \label{varphi_chi-SS3}
\eea where we have kept $f_{,\phi\phi}$ term compared with $f_{,X}
k^2/a^2$ term, and order $1$ compared with $(F_{,R}/F) k^2/a^2$ [see
below Eq.\ (\ref{alpha_chi-2})] as in
\cite{Tsujikawa-2007,Tsujikawa-etal-2008,DeFelice-etal-2010}; we
{\it ignored} $f_{,X \phi} \delta X$ term in an expansion of $\delta
f_{,\phi}$ term in Eq.\ (\ref{EOM-pert-f}). From Eqs.\
(\ref{varphi_chi-SS1})-(\ref{varphi_chi-SS3}), we have \bea
   \varphi_\chi
       \mskip-6mu&\simeq&\mskip-6mu
       - { \left( 1 + 2 {F_{,R} \over F} {k^2 \over a^2} \right)
       \left( f_{,\phi\phi}
       + f_{,X} {k^2 \over a^2} \right)
       - 2 {( F_{,\phi} )^2 \over F} {k^2 \over a^2}
       \over
       \left( 1 + 4 {F_{,R} \over F} {k^2 \over a^2} \right)
       \left( f_{,\phi\phi}
       + f_{,X} {k^2 \over a^2} \right)
       - 4 {( F_{,\phi} )^2 \over F} {k^2 \over a^2 } }
   \nonumber \\
   & & \times
       \alpha_\chi
       \equiv - \bar \eta \alpha_\chi,
   \label{varphi_chi-general} \\
   {k^2 \over a^2} \alpha_\chi
       \mskip-6mu&\simeq&\mskip-6mu - {1 \over 2 F}
       { \left( 1 + 4 {F_{,R} \over F} {k^2 \over a^2} \right)
       \left( f_{,\phi\phi}
       + f_{,X} {k^2 \over a^2} \right)
       - 4 {( F_{,\phi} )^2 \over F} {k^2 \over a^2}
       \over
       \left( 1 + 3 {F_{,R} \over F} {k^2 \over a^2} \right)
       \left( f_{,\phi\phi}
       + f_{,X} {k^2 \over a^2} \right)
       - 3 {( F_{,\phi} )^2 \over F} {k^2 \over a^2 } }
   \nonumber \\
   & & \times
       \delta \mu_{c v_c}.
   \label{alpha_chi-general}
\eea These can be compared with Eqs.\ (\ref{mod-1}) and
(\ref{mod-2}). However, it is important to notice that $\delta
\mu_{c v_c}$ differs from $\delta \mu_v$. By comparing Eqs.\
(\ref{Poisson-eq}) and (\ref{varphi_chi-SS2}), and using Eqs.\
(\ref{varphi_chi-SS1}) and (\ref{varphi_chi-general}), we have \bea
   & & \delta \mu_v
       \simeq {2 \bar \eta \over 1 + \bar \eta} {1 \over F}
       \delta \mu_{c v_c}.
   \label{delta-mu_v-small}
\eea Using this, we have \bea
   {k^2 \over a^2} \alpha_\chi
   &\simeq& - {1 \over 1 + \bar \eta} {1 \over F} \delta \mu_{c v_c}
   \nonumber \\
   &\simeq& - {1 \over 2 \bar \eta} \delta \mu_v
       \equiv - {1 \over 2} \bar \mu \delta \mu_v.
   \label{alpha_chi-general-2}
\eea Therefore, we have \bea
   & & \bar \mu \simeq {1 \over \bar \eta},
\eea which is an already presented result in an exact context in
Eq.\ (\ref{Pi-alpha_chi}). The above relations are derived in the
general context of $f(R, \phi, X)$ gravity which usually leads to
higher order theory: see below Eq.\ (\ref{action}).

From Eqs.\ (\ref{ddot-delta-eq-CCG-SS}) and
(\ref{alpha_chi-general-2}) we have \bea
   \ddot \delta_{c v_c} + 2 H \dot \delta_{c v_c}
   &\simeq& {1 \over 2} \bar \mu \delta \mu_v
       \simeq {1 \over 2} {2 \bar \mu
       \over 1 + \bar \mu} {1 \over F} \delta \mu_{c v_c}
   \nonumber \\
   &\equiv& 4 \pi G_{\rm eff} \delta \mu_{c v_c},
   \label{ddot-delta_c-eq}
\eea where the effective gravitational constant becomes \bea
   & & 8 \pi G_{\rm eff}
       \simeq {2 \bar \mu
       \over 1 + \bar \mu} {1 \over F}.
   \label{Geff}
\eea For $f = R - 2 X - 2 V$ we have a correct Einstein's gravity
result with $8 \pi G_{\rm eff} = 1$.

Now, we consider two cases: (i) $f = R F(\phi) + 2 p(\phi, X)$
theory, and (ii) $f = f(R)$ theory.

(i) In the case of $f = R F(\phi) + 2 p(\phi, X)$ theory, Eqs.\
(\ref{varphi_chi-general}) and (\ref{alpha_chi-general}) reduce to
\bea
   \varphi_\chi
   &\simeq&
       - { \left( f_{,X} - 2 {(F_{,\phi})^2 \over F} \right) {k^2 \over a^2}
       + f_{,\phi\phi} \over
       \left( f_{,X} - 4 {(F_{,\phi})^2 \over F} \right) {k^2 \over a^2}
       + f_{,\phi\phi} } \alpha_\chi,
   \\
   {k^2 \over a^2} \alpha_\chi
   &\simeq& - {1 \over 2F}
       { \left( f_{,X} - 4 {(F_{,\phi})^2 \over F} \right) {k^2 \over a^2}
       + f_{,\phi\phi} \over
       \left( f_{,X} - 3 {(F_{,\phi})^2 \over F} \right) {k^2 \over a^2}
       + f_{,\phi\phi} } \delta \mu_{c v_c}.
   \label{alpha_chi-1}
\eea

(ii) In the case of $f = f(R)$ theory, however, the equation of
motion in Eq.\ (\ref{EOM-pert-f}) is not available. Instead, from
Eq.\ (\ref{delta-R}) we have \bea
   \delta R_\chi
       \simeq 2 {k^2 \over a^2}
       \left( \alpha_\chi + 2 \varphi_\chi \right)
       \simeq {1 \over F} \delta \mu_{c \chi}
       - 3 {k^2 \over a^2} {\delta F_\chi \over F}.
   \label{delta-R-SS}
\eea We take $\delta_{c \chi} \simeq \delta_{c v_c}$ in the
small-scale limit. From Eqs.\ (\ref{delta-R-SS}) and
(\ref{varphi_chi-SS1}) we recover Eq.\ (\ref{varphi_chi-SS2}). Now,
in the $f(R)$ theory we have $\delta F = F_{,R} \delta R$. Using
this relation, from Eqs.\ (\ref{varphi_chi-SS1}) and
(\ref{varphi_chi-SS2}) together with Eq.\ (\ref{delta-R-SS}) we can
derive \bea
   \varphi_\chi
   &\simeq&
       - {1 + 2 {F_{,R} \over F} {k^2 \over a^2} \over
       1 + 4 {F_{,R} \over F} {k^2 \over a^2}}
       \alpha_\chi,
   \\
   {k^2 \over a^2} \alpha_\chi
   &\simeq& - {1 \over 2F}
       {1 + 4 {F_{,R} \over F} {k^2 \over a^2} \over
       1 + 3 {F_{,R} \over F} {k^2 \over a^2}}
       \delta \mu_{c v_c}.
   \label{alpha_chi-2}
\eea Although these results can be regarded as cases of Eqs.\
(\ref{varphi_chi-general}) and (\ref{alpha_chi-general}), we need a
separate treatment because Eqs.\ (\ref{varphi_chi-general}) and
(\ref{alpha_chi-general}) were derived from the equation of motion
which is not available in $f(R)$ gravity. As in Eqs.\
(\ref{varphi_chi-SS3}) and (\ref{alpha_chi-general}) we kept $\delta
R = \delta F/F_{,R}$ term even compared with $(k^2/a^2) \delta F/F$
term because $F_{,R}$ becomes zero in the Einstein gravity limit.
Thus, it is important to notice that even in the small-scale limit
we are already considering, we have to set $F_{,R}/F \ll a^2/k^2 \ll
1$ to properly have the Einstein's gravity limit.

\begin{figure*}
\begin{center}
\includegraphics[width=8.5cm]{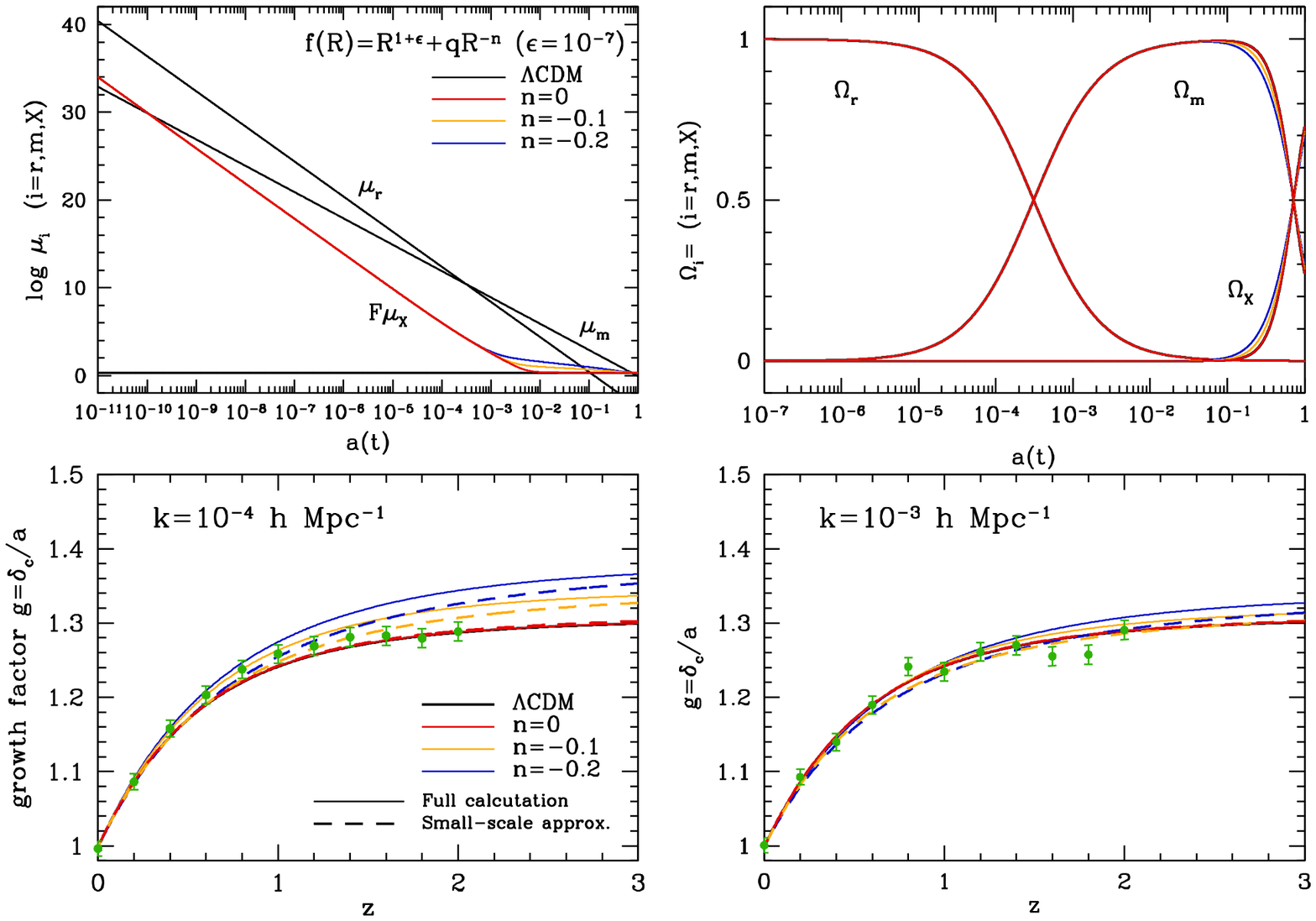}
\includegraphics[width=8.5cm]{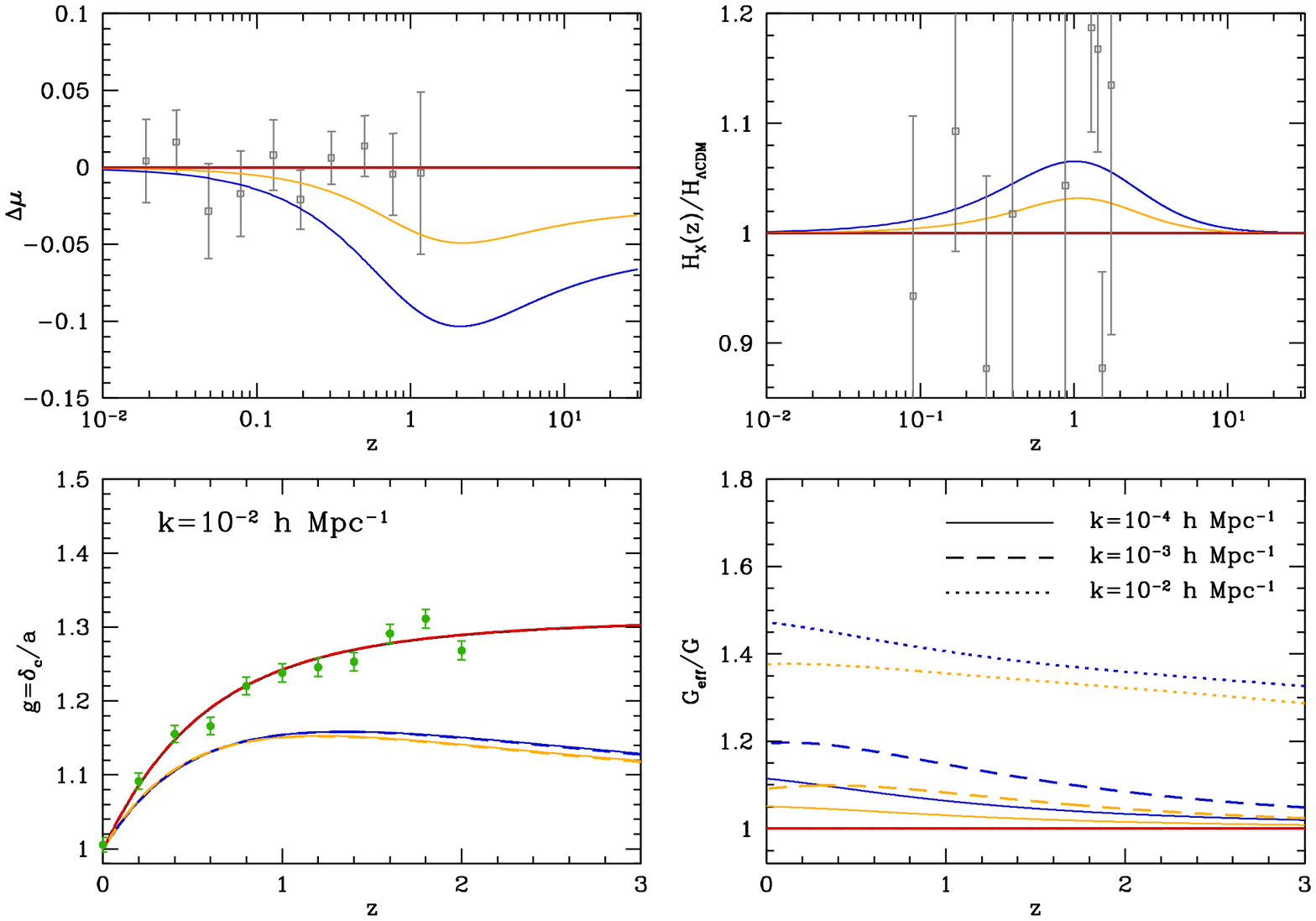}
\caption{
         Top panels from left to right:
         Evolution of density $\mu_i$ ($i = r, m, X$);
         recent behavior of density parameters due to $f(R)$ gravity;
         test under the type-Ia supernovae observation;
         evolution of the Hubble parameter.
         We use a functional form
         $f(R)=R^{1+\epsilon}+qR^{-n}$ with
         $\epsilon=10^{-7}$ and $n=0$ (red), $-0.1$ (yellow), $-0.2$
         (blue curves).
         Note that we compare $\mu_r$ and $\mu_m$ with $F\mu_X$.
         Evolution of $F \mu_X$ shows early scaling
         and its recent domination as a dark energy.
         We use definitions,
         $\Omega_i \equiv \mu_i/(3FH^2)$ for radiation and matter
         and $\Omega_X\equiv \mu_X/(3H^2)$.
         The binned type-Ia supernovae data are based on the Union2 sample
         \cite{Amanullah-etal-2010}, and Hubble parameteter data are from Ref.\
         \cite{Samushia-Ratra-2006}.
         Bottom panels from left to right:
         Behaviors of the CDM density perturbation growth factor in
         three different scales; the last figure shows
         evolutions of $G_{\rm eff}$.
         In the growth factor panels, results from the full calculation
         (including radiation; solid curves) and from the small
         scale approximation (dashed curves) are shown together.
         Green dots with $1$\% error bars are expectations
         from the future X-ray and weak lensing observations
         \cite{Vikhlinin-etal-2009}.
         As the scale becomes smaller results from the small-scale approximation
         more properly reproduce ones from full calculation.
         The evolutions of $G_{\rm eff}$, however,
         show that deviation of $G_{\rm eff}$ from Newtonian $G$
         is more significant as the scale becomes smaller.}
\label{fig:comparison}
\end{center}
\end{figure*}

Apparently, the derivation of Eq.\ (\ref{ddot-delta_c-eq}) in the
small-scale limit requires rather complicated steps, and the only
way to check its validity is comparing with the exact result. In
Fig.\ \ref{fig:comparison} we estimate the accuracy of the above
asymptotic approximations using a case of $f(R)$ gravity. We
consider a case with $f = R^{1+\epsilon} + q R^{-n}$. The $R^{1 +
\epsilon}$ term dominates in the early (radiation) era and allows a
scaling evolution of $\mu_X$ and $\delta \mu_X$ following the
dominant fluid; first follows radiation and then matter. The $q
R^{-n}$ term can be tuned to cause accelerated evolution in recent
era; we have studied this case in \cite{fR-2010} in details. For the
perturbations we solve equations in the CDM-comoving gauge presented
in Eqs.\ (\ref{fR-dot-F-eq})-(\ref{fR-dot-CDM-eq}) with additional
presence of radiation component.

The Figure shows that in the small
scale far inside the horizon the approximate equations in Eqs.\
(\ref{ddot-delta-eq-CCG-SS}) and (\ref{alpha_chi-2}) reproduce the
evolution of $\delta_{c v_c}$ properly.
         That is, as the scale becomes smaller results from the small-scale approximation
         more properly reproduce ones from full calculation.
         The evolutions of $G_{\rm eff}$, however,
         show that deviation of $G_{\rm eff}$ from Newtonian $G$
         is more significant as the scale becomes smaller, see Eq.\ (\ref{alpha_chi-2}).
In practical applications it is always important to check the validity of such
asymptotic form approximations compared with exact solutions based
on the complete set of equations. The complete set of perturbation
equations in several gauges in our generalized gravity theories is
presented in the Appendix C.

\section{Potential cosmological signatures of non-Einstein gravity}
                                             \label{sec:loopholes}

Can we find evidence of non-Einstein gravity nature from the large
scale cosmological observations? Before presenting a (in principle)
possible way of using cosmological observations to distinguish
potential non-Einstein' gravity nature, here, we would like to point
out some potential loopholes in recent such arguments based on
pseudo-Newtonian approaches motivated by approximate treatment of
modified gravity theories.

One argument is based on the deviation of baryon (or CDM) density
perturbation equation and Poisson's equation in modified gravity. In
the small-scale limit from Eqs.\ (\ref{ddot-delta-eq-CCG-SS}),
(\ref{ddot-delta_c-eq}) and (\ref{Geff}) we have \bea
   & & \ddot \delta_{c v_c} + 2 H \dot \delta_{c v_c}
       \simeq - {k^2 \over a^2} \alpha_\chi,
   \\
   & & {k^2 \over a^2} \alpha_\chi
       \simeq - 4 \pi G_{{\rm eff}} \delta \mu_{c v_c}.
   \label{Poisson-eq-proper}
\eea Under approximations used in previous section we have Eq.\
(\ref{Geff}), and in the Einstein's gravity limit we have $8 \pi
G_{\rm eff} = 1$. Here, it is important to remember that the closed
form second-order equation in the considered modified gravity
theories is valid only in the small-scale limit.

Although in Fig.\ \ref{fig:comparison} we have shown an example
where the CDM (or baryon) growth rate is well approximated by the
small-scale limit approximation in Eq.\ (\ref{ddot-delta_c-eq}),
that is not necessarily the case for other cosmological observations
like the CMB and the density power spectra, etc. It is important to
notice that even in Einstein's gravity, in the presence of dynamic
dark energy, in general, we cannot ignore the contribution of the
dark energy perturbation in the right-hand-side of the Poisson's
equation: see Eqs.\ (\ref{effective-fluid-Bean}) and
(\ref{delta-mu_v}). One such an example demonstrating the importance
of coupling even in Einstein's gravity is presented in
\cite{Park-etal-PRL-2009}.

In \cite{Park-etal-PRL-2009} we have shown that even in the case of
Einstein's gravity we have observationally significant deviations by
ignoring the dark energy perturbations in the CMB and density power
spectra, and in the baryon density perturbation growth rate. As we
emphasized in this work, consideration of dynamical dark energy in
general causes a coupling between the baryon (or CDM) density
perturbation and the dark energy perturbation; and the dynamics of
dark energy perturbation is described by its own additional equation
of motion which is again coupled through gravity with other
components. This is true even in the context of Einstein's gravity
with a minimally coupled scalar field as a dynamical dark energy.

\begin{figure}
\begin{center}
\includegraphics[width=8.6cm]{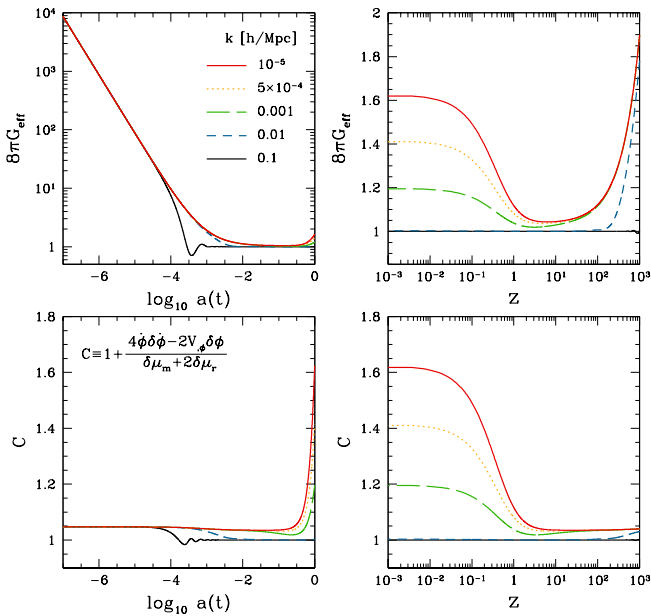}
\caption{
         Top panels: Evolution of $8 \pi G_{{\rm eff}}$ as a function
         of scale factor $a(t)$ and the redshift $z$
         for several different scales.
         As the scale becomes smaller $8 \pi G_{{\rm eff}}$
         approaches unity.
         We consider a double exponential potential $ V(\phi) = V_1 e^{- \lambda_1 \phi}
         + V_2 e^{- \lambda_2 \phi} $ where $\phi$ is the scalar
         field and $\lambda_2 = 1.0$;
         for other parameters, see \cite{Park-etal-PRL-2009}.
         Bottom panels: Evolution of
         the ratio of the scalar field compared with
         the ordinary fluid (matter plus radiation) using
         the same model as in the top panel.
         Deviations of the ratio from unity in the early radiation
         and matter dominated eras are due to the scaling nature
         of scalar field we considered.
         }
\label{fig:MSF}
\end{center}
\end{figure}

In \cite{Park-etal-PRL-2009} we have studied a case of dark energy
model based on a minimally coupled scalar field with a double
exponential potential as an example. We showed that proper
consideration of the dark energy perturbations is important in
comparing theories with present day cosmological observations. In
the CMB temperature anisotropy power spectrum, the large-scale
density power spectrum, and even in the perturbation growth rate we
have shown that it is critically important to take into account of
dark energy perturbations properly. This result can be regarded as
an evidence that the coupling between the baryon (or CDM) density
perturbation and the dynamical dark energy perturbation is in
general important even in Einstein's gravity. In the presence of a
scalar field dark energy, in the CCG, from Eqs.\ (\ref{eq-5}),
(\ref{eq-6i}), (\ref{eq-7i}) and (\ref{fluid-X}) we have \bea
   \ddot \delta_c + 2 H \dot \delta_c
   &=& {1 \over 2} \left( \delta \mu + 3 \delta p \right)
   \nonumber \\
   &=& {1 \over 2} \left( \delta \mu_c
       + 2 \delta \mu_r
       + 4 \dot \phi \delta \dot \phi
       - 2 V_{,\phi} \delta \phi \right)
   \nonumber \\
   &\equiv& 4 \pi G_{\rm eff} \delta \mu_c.
\eea In Fig.\ \ref{fig:MSF} we present evolution of $8 \pi G_{{\rm
eff}}$ and the effect of the scalar field contribution using a
double exponential potential model of a minimally coupled scalar
field in Einstein's gravity \cite{Park-etal-PRL-2009}. Although
becomes negligible as we go to the small scale, we have $8 \pi
G_{{\rm eff}} \neq 1$ even for a minimally coupled scalar field as
the dark energy in Einstein's gravity.

%
%

The other argument is based on Eq.\ (\ref{mod-1}) so that $\bar \eta
\neq 1$ could be regarded as a strong signature of the non-Einstein
gravity nature. The main point used in this argument is that the
gravitational weak lensing effect can measure the {\it sum} of
Newtonian potential ($- \alpha_\chi$) and the post-Newtonian
potential ($\varphi_\chi$). From Eq.\ (\ref{eq-4}) we have \bea
   & & \alpha_\chi + \varphi_\chi = - \Pi,
   \label{weak-lensing}
\eea Although we could have significant amount of anisotropic stress
$\Pi$ even in Einstein's gravity, that could be indeed regarded as
highly peculiar situation almost comparable to rather considering
the non-Einstein gravity. As we have $\varphi_\chi = - \alpha_\chi$
in Einstein's gravity, we can accept that it is reasonable to regard
$\Pi = ( \bar \eta - 1 ) \alpha_\chi \neq 0$, thus $\bar \eta \neq
1$ as a strong signature of non-Einstein's gravity. Now, as
$\alpha_\chi - \varphi_\chi$ can be measured from the weak lensing
observation, the remaining problem is whether we can determine
$\alpha_\chi$ or $\varphi_\chi$ individually in Einstein's gravity.

Here, it is interesting to point out that the evolution of baryonic
velocity perturbation $v_{b \chi}$, in principle, can determine the
Newtonian potential $\alpha_\chi$
\cite{Song-Dore-2009,Song_etal-2010-11}. Independently of the
gravity theories, from Eq.\ (\ref{eq-7i}) or (\ref{eq-7m}) for the
baryon component, we have \bea
   & & \dot v_{b \chi} + H v_{b \chi}
       = {k \over a} \alpha_\chi.
   \label{baryon-velocity-eq}
\eea Similar equation for the CDM component is presented in Eq.\
(\ref{Bean-eq2}). Thus, apparently, the velocity perturbation
depends only on the Newtonian gravitational potential $\alpha_\chi$.
Since this equation is derived from the momentum conservation of the
baryon component, unless there exist exotic interaction with other
components, this equation is generally valid independently of the
nature of the gravity theory based on Riemannian geometry.

Therefore, in Einstein's gravity theory we have \bea
   & & \alpha_\chi - \varphi_\chi = 2 \alpha_\chi.
   \label{WL-velocity}
\eea In this expression the left-hand-side and the right-hand-side
can be determined from weak lensing observation and peculiar
velocity field observation, respectively. By combining these two
informations we can, in principle, tell whether the post-Newtonian
gravitational potential ($\varphi_\chi$) has the same amplitude as
the Newtonian potential ($- \alpha_\chi$). If the two amplitudes
differ from each other, i.e. if Eq.\ (\ref{WL-velocity}) is
violated, assuming validity of linear theory, that can be regarded
as an important signature of non-Einstein's gravity nature revealed
in the cosmological situation. For recent study in this direction,
see \cite{Song_etal-2010-11}.

\section{Discussion}

We have considered feasibility of modifying Newton's gravity in
cosmology from the perspective of generalized gravity theories. As a
concrete example we examined $f(R, \phi, X)$ gravity which includes
$f(R)$ gravity and many other gravity theories known in the
literature as cases. For the baryon or CDM component in the context
of the generalized gravity theory, the perturbation equations are
inevitably coupled with the presence of the generalized gravity
theory sector, thus in general ending up with at least fourth-order
differential equation; situation is the same even in the minimally
coupled scalar field as a dark energy in Einstein's gravity. The
closed form of CDM density perturbation equation proposed in the
literature is only possible in the small-scale limit in the context
of generalized relativistic gravity theories. Even in such a
small-scale limit we can easily introduce cases where proper
treatment of generalized gravity is important to compare with the
CMB and density power spectra.

It is always important to analyze properly (i.e., without
approximation) the roles of dynamic dark energy perturbation in
Einstein's gravity before one concludes against the canonical
gravity theory which has been observationally successful in all weak
gravity tests. We also have addressed potential loopholes in arguing
non-Einstein gravity nature based on cosmologically modified
pseudo-Newtonian gravity theories. We point out that future precise
observations of the weak lensing together with the baryon velocity
perturbation can potentially test non-Einstein gravity nature based
on cosmological observations.

%
%
\section*{Appendix}

\section*{Appendix A: Perturbations in relativistic gravity}

In the Appendix A and B we consider the presence of general $K$.
Background evolution is described by \bea
   & & H^2 = {1 \over 3} \mu - {K \over a^2}, \quad
       \dot \mu = - 3 H \left( \mu + p \right).
   \label{BG-eqs}
\eea In the multiple component case, we have $\mu = \sum_j \mu_j$
and $p = \sum_j p_j$, together with \bea
   & & \dot \mu_i
       = - 3 H ( \mu_i + p_i ) + I_{0i},
   \label{BG-mu_i}
\eea where $\sum_j I_{0j} \equiv 0$. Based on our metric and
energy-momentum tensor conventions in Eqs.\ (\ref{metric}) and
(\ref{Tab}), the scalar-type perturbations are described by
\cite{Bardeen-1988} \bea
   & & \kappa \equiv - 3 \dot \varphi + 3 H \alpha  + {k^2 \over a^2} \chi,
   \label{eq-1} \\
   & & {1 \over 2} \delta \mu + H \kappa
       = {k^2 - 3K \over a^2} \varphi,
   \label{eq-2} \\
   & & \kappa
       - {k^2 - 3K \over a^2} \chi
       = {3 \over 2} (\mu + p) {a \over k} v,
   \label{eq-3} \\
   & & \dot \chi + H \chi - \varphi - \alpha
       = \Pi,
   \label{eq-4} \\
   & & \dot \kappa + 2 H \kappa
       + \left( 3 \dot H - {k^2 \over a^2} \right) \alpha
       = {1 \over 2} \left( \delta \mu + 3 \delta p \right),
   \label{eq-5} \\
   & & \delta \dot \mu + 3 H \left( \delta \mu + \delta p \right)
       = \left( \mu + p \right) \left( \kappa - 3 H \alpha
       - {k \over a} v \right),
   \label{eq-6} \\
   & & {[a^4 (\mu + p) v]^{\displaystyle\cdot} \over a^4(\mu + p)}
       = {k \over a} \alpha
       + {k \over a (\mu + p)} \left( \delta p
       - {2 \over 3} {k^2 - 3K \over a^2} \Pi \right),
   \nonumber \\
   \label{eq-7}
\eea where $\chi \equiv a \beta + a^2 \dot \gamma$ is a spatially
gauge-invariant combination. We note that all our perturbation
equations are spatially gauge invariant. Although derived in
Einstein's gravity these equations are valid even in the context of
generalized gravity theories including the $f(R, \phi, X)$ gravity
presented in the Appendix B. The generalized gravity theory we
consider can be written in the Einstein's gravity form \bea
   & & G_{ab} = T_{ab},
   \label{G_ab}
\eea where $T_{ab}$ is reinterpreted as the effective
energy-momentum tensor \cite{HN-GGT-2005}. In this way, as long as the
generalized gravity can be expressed as in Eq.\ (\ref{G_ab}), the
above basic set of perturbation equations is valid in those gravity
theories as well.

In the presence of multiple component fluids (including dust, dark
matter, radiation, fields, etc.), we have \bea
   & & \delta \mu = \sum_j \delta \mu_j, \quad
       \delta p = \sum_j \delta p_j,
   \nonumber \\
   & &
       (\mu + p) v = \sum_j (\mu_j + p_j) v_j, \quad
       \Pi = \sum_j \Pi_j.
   \label{fluid-sum}
\eea In the case of generalized gravity theory, see Eqs.\
(\ref{effective-fluid}) and (\ref{fluid-sum-X}). The individual
component follows its own equation of motion
\begin{widetext}
\bea
   & & \delta \dot \mu_i + 3 H \left( \delta \mu_i + \delta p_i \right)
       = \left( \mu_i + p_i \right) \left( \kappa - 3 H \alpha
       - {k \over a} v_i \right)
       - {1 \over a} \delta I_{0i},
   \label{eq-6i} \\
   & & {[a^4 (\mu_i + p_i) v_i]^{\displaystyle\cdot} \over a^4(\mu_i + p_i)}
       = {k \over a} \alpha
       + {k \over a (\mu_i + p_i)} \left( \delta p_i
       - {2 \over 3} {k^2 - 3K \over a^2} \Pi_i
       - \delta I_{i} \right),
   \label{eq-7i}
\eea where $\sum_j \delta I_{0j} \equiv 0 \equiv \sum_j \delta I_j$.

\end{widetext}

%
%
\section*{Appendix B: Perturbations in $f(R,\phi,X)$ gravity}

We consider $f(R,\phi,X)$ gravity considered in \cite{HN-GGT-2005}.
The action is \bea
   & & S = \int d^4 x \sqrt{-g} \left[
       {1 \over 2} f (R, \phi, X) + L_{m} \right],
   \label{action}
\eea where $X \equiv {1 \over 2} \phi^{,c} \phi_{,c}$. The minimally
coupled scalar field is a case with $f = R - 2 X - 2 V(\phi)$. Even
in the absence of additional matter component the $f (R, \phi, X)$
gravity, in general, leads to a fourth-order differential equation
for the scalar-type perturbations. In order to have second-order
differential equation with nontrivial $F \equiv {\partial f/\partial
R}$ we should have either (i) $f = R F(\phi, X) + 2 p(\phi, X)$ or
(ii) $f = f(R)$, see \cite{HN-GGT-2005}.

The gravitational field equation following from the above action can
be arranged in the form of Einstein's gravity in which the new
contributions are interpreted as effective energy-momentum tensor;
the effective fluid quantities are introduced in Eq.\ (\ref{Tab}).
Using this strategy, the background and perturbation equations in
Einstein's gravity remain valid with the fluid quantities replaced
by the effective ones \cite{HN-GGT-1996}.

Background evolution is still described by Eq.\ (\ref{BG-eqs}) with
the fluid quantities reinterpreted as the following effective ones
\bea
   & & \mu = {1 \over F} \mu_m + \mu_X, \quad
       p = {1 \over F} p_m + p_X,
   \nonumber \\
   & &
       \mu_X \equiv {1 \over F} \left( f_{,X} X
       + {FR - f \over 2} - 3 H \dot F \right),
   \nonumber \\
   & &
       p_X \equiv {1 \over F} \left( - {FR - f \over 2}
       + \ddot F + 2 H \dot F \right),
   \label{fluid-BG}
\eea where $X \equiv - {1 \over 2} \dot \phi^2$. The equations of
motion is presented in Eq.\ (80) of \cite{HN-GGT-2005} \bea
   & & {1 \over a^3} \left( a^3 \dot \phi f_{,X} \right)^{\displaystyle\cdot}
       + f_{,\phi} = 0.
   \label{EOM-BG-f}
\eea We also have \bea
   & & R = 6 \left( 2 H^2 + \dot H + {K \over a^2} \right)
       = \mu - 3 p.
   \label{R}
\eea The ordinary matter part follows its own equation of motion
$\dot \mu_m = - 3 H \left( \mu_m + p_m \right)$. Thus, we have \bea
   & & \dot \mu_X + 3 H \left( \mu_X + p_X \right)
       = {\dot F \over F^2} \mu_m.
\eea 

\begin{widetext}
In our generalized gravity context Eqs.\ (\ref{eq-1})-(\ref{eq-7})
remain valid with the fluid quantities reinterpreted as the
following effective ones
 \bea
   & & \delta \mu = {1 \over F} \delta \mu_m
       + \delta \mu_X - \mu_m {\delta F \over F^2}, \quad
       \delta p = {1 \over F} \delta p_m
       + \delta p_X - p_m {\delta F \over F^2},
   \nonumber \\
   & & \left( \mu + p \right) v
       = {1 \over F} \left( \mu_m + p_m \right) v_m
       + \left( \mu_X + p_X \right) v_X, \quad
       \Pi = {1 \over F} \Pi_m + \Pi_X,
   \label{effective-fluid}
\eea with \bea
   & & \delta \mu_X
       = {1 \over F} \Bigg[ {1 \over 2}
       \left( f_{,X} \delta X - f_{,\phi} \delta \phi \right)
       + \delta f_{,X} X
       - 3 H \delta \dot F
       + \left( {1 \over 2} f - f_{,X} X + 3 H \dot F
       - F {k^2 \over a^2} \right) {\delta F \over F}
       + \dot F \kappa
       + 3 H \dot F \alpha \Bigg],
   \nonumber \\
   & & \delta p_X
       = {1 \over F} \Bigg[ {1 \over 2}
       \left( f_{,X} \delta X
       + f_{,\phi} \delta \phi \right)
       + \delta \ddot F
       + 2 H \delta \dot F
       - \left( {1 \over 2} f + \ddot F + 2 H \dot F
       - F {2 \over 3} {k^2 \over a^2} \right) {\delta F \over F}
       - {2 \over 3} \dot F \kappa
       - \dot F \dot \alpha
       - 2 \left( \ddot F + H \dot F \right) \alpha \Bigg],
   \nonumber \\
   & & \left( \mu_X + p_X \right) v_X
       = - {k\over aF}
       \left( {1 \over 2} f_{,X} \dot \phi \delta \phi
       - \delta \dot F
       + H \delta F
       + \dot F \alpha \right),
   \nonumber \\
   & & \Pi_X
       = {1 \over F} \left( \delta F - \dot F \chi \right),
   \label{fluid-X}
\eea and $\delta X = - \dot \phi \delta \dot \phi + \dot \phi^2
\alpha$. If we have multiple fluid components, we have \bea
   & & \mu_m = \sum_j \mu_j, \quad
       p_m = \sum_j p_j,
   \nonumber \\
   & &
       \delta \mu_m = \sum_j \delta \mu_j, \quad
       \delta p_m = \sum_j \delta p_j, \quad
       \left( \mu_m + p_m \right) v_m
       = \sum_j \left( \mu_j + p_j \right) v_j, \quad
       \Pi_m = \sum_j \Pi_j,
   \label{fluid-sum-X}
\eea where the individual component satisfies Eq.\ (\ref{BG-mu_i})
for the background and Eqs.\ (\ref{eq-6i}) and (\ref{eq-7i}) for the
perturbations. The equation of motion is presented in Eq.\ (81) of
\cite{HN-GGT-2005} \bea
   & & f_{,X} \left[ \delta \ddot \phi
       + \left( 3 H + {(f_{,X})^{\displaystyle\cdot} \over f_{,X}} \right)
       \delta \dot \phi
       + {k^2 \over a^2} \delta \phi
       + \dot \phi \left( 3 \dot \varphi - \dot \alpha
       - {k^2 \over a^2} \chi \right) \right]
       + 2 f_{,\phi} \alpha
       + {1 \over a^3} \left( a^3 \dot \phi \delta f_{,X}
       \right)^{\displaystyle\cdot}
       + \delta f_{,\phi} = 0.
   \label{EOM-pert-f}
\eea We also have \bea
   & & \delta R = 2 \left[ - \dot \kappa
       - 4 H \kappa
       + \left( {k^2 \over a^2} - 3 \dot H \right) \alpha
       + 2 {k^2 - 3 K \over a^2} \varphi \right]
       = \delta \mu - 3 \delta p.
   \label{delta-R}
\eea If the ordinary matter part follows its own equation of motion
\bea
   & & \delta \dot \mu_m + 3 H \left( \delta \mu_m + \delta p_m \right)
       = \left( \mu_m + p_m \right) \left( \kappa - 3 H \alpha
       - {k \over a} v_m \right),
   \label{eq-6m} \\
   & & {[a^4 (\mu_m + p_m) v_m]^{\displaystyle\cdot} \over a^4(\mu_m + p_m)}
       = {k \over a} \alpha
       + {k \over a (\mu_m + p_m)} \left( \delta p_m
       - {2 \over 3} {k^2 - 3K \over a^2} \Pi_m \right),
   \label{eq-7m}
\eea which follow from Eqs.\ (\ref{eq-6i}) and (\ref{eq-7i}), from
Eqs.\ (\ref{eq-6}) and (\ref{eq-7}) we have \bea
   & & \delta \dot \mu_X
       + 3 H \left( \delta \mu_X + \delta p_X \right)
       - \left( \mu_X + p_X \right)
       \left( \kappa - 3 H \alpha
       - {k \over a} v_X \right)
       = {\dot F \over F^2} \delta \mu_m
       + \mu_m {1 \over F^2} \left( \delta \dot F
       - 2 {\dot F \over F} \delta F \right),
   \label{eq-6X} \\
   & & {1 \over a^4} \left[ a^4 \left( \mu_X + p_X \right) v_X
       \right]^{\displaystyle\cdot}
       - {k \over a} \left[ \left( \mu_X + p_X \right) \alpha
       + \delta p_X
       - {2 \over 3} {k^2 - 3 K \over a^2} \Pi_X \right]
       = {\dot F \over F^2} \left( \mu_m + p_m \right) v_m
       - {1 \over F^2} {k \over a} p_m \delta F.
   \label{eq-7X}
\eea Notice that in the single component case without $m$-component,
Eqs.\ (\ref{eq-6X}) and (\ref{eq-7X}) are the same as Eqs.\
(\ref{eq-6m}) and (\ref{eq-7m}) with the sub-indices $m$ replaced by
$X$.

\section*{Appendix C: CDM perturbation in $f(R, \phi, X)$ gravity}

Here we set $K = 0$. For simplicity, we consider (i) $f = R F(\phi)
+ 2 p(\phi, X)$ and (ii) $f = f(R)$. In the following, we present
complete sets of equations of CDM in the above generalized gravity
theories, in three different gauge conditions. These are the
zero-shear gauge, the uniform-field (or $F$) gauge, and the
CDM-comoving gauge.

\subsection*{Appendix C.1: Zero-shear gauge}

The zero-shear gauge takes $\chi \equiv 0$ as the temporal gauge
condition. Equations (\ref{Bean-eq1}) and (\ref{Bean-eq2}) are valid
for differential equations for $\delta_{c \chi}$ and $v_{c \chi}$.
Using Eq.\ (\ref{varphi-alpha-Bean}) $\alpha_\chi$ can be expressed
in terms of $\varphi_\chi$ and $\delta F_\chi$. Thus, two additional
first-order differential equations for $\varphi_\chi$ and $\delta
F_\chi$ will complete the perturbation equations. One equation
follows from Eqs.\ (\ref{eq-1}) and (\ref{eq-3}) \bea
   & & \dot \varphi
       + \left( H + {\dot F \over 2F} \right) \varphi
       = {1 \over 2 F} \Bigg[
       {1 \over 2} f_{,X} \dot \phi \delta \phi
       - \delta \dot F
       - {1 \over 2} \left( H + {\dot F \over F} \right)
       {\delta F \over F}
       - \mu_c {a \over k} v_{c} \Bigg].
   \label{Bean-eq3}
\eea The other one can be derived from Eq.\ (\ref{Poisson-eq}) \bea
   & & {1 \over 2} \left( f_{,X} + 2 f_{,XX} X \right)
       \dot \phi \delta \dot \phi
       - {3 \dot F \over 2 F} \delta \dot F
       + {1 \over 2} \left[ f_{,\phi}
       - 2 f_{,\phi X} X
       + 3 \left( H + {\dot F \over 2F} \right) f_{,X} \dot \phi
       \right] \delta \phi
   \nonumber \\
   & & \quad
       + \left[ 3 H^2 + {\mu_c \over F}
       - {3 \dot F \over 2F} H
       - {1 \over F} \left(
       {3 \dot F^2 \over 2F}
       + {1 \over 2} f
       + 2 f_{,XX} X^2 \right)
       + {k^2 \over a^2} \right] \delta F
   \nonumber \\
   & & \quad
       = \left(
       {3 \dot F^2 \over 2F}
       + f_{,X} X
       + 2 f_{,XX} X^2
       - 2 F {k^2 \over a^2} \right) \varphi
       + \delta \mu_c
       + 3 \left( H + {\dot F \over 2 F} \right) {a \over k} \mu_c v_c,
   \label{Bean-eq4}
\eea where we have $\delta F = F_{,\phi} \delta \phi$ for (i), and
$\delta \phi = 0 = X$ for (ii). From these we have the two equations
for $\dot \varphi_\chi$ and $\delta \dot F_\chi$. And these two
equations are the additional equations we need to solve together
with Eqs.\ (\ref{Bean-eq1}) and (\ref{Bean-eq2}).
\end{widetext}

\subsection*{Appendix C.2: Uniform-field gauge}

We take the uniform-field gauge for (i), and the uniform-$F$ gauge
for (ii); in both cases we set $\delta \phi = 0 = \delta F$ as the
temporal gauge condition, and we call the gauge condition as UFG. We
have $\delta X = - 2 X \alpha$ and $\delta f_{,X} = - 2 f_{,XX} X
\alpha$. From Eqs.\ (\ref{eq-1})-(\ref{eq-3}), Eqs.\ (\ref{eq-1}),
(\ref{eq-3}), and (\ref{eq-4}), respectively, we have \bea
   & & \dot \Phi
       = {\dot F + 2 H F \over {3 \dot F^2 \over 2 F}
       + f_{,X} X + 2 f_{,XX} X^2}
       \Bigg[ - {k^2 \over a^2} \Psi
       + {1 \over 2F} \delta \mu_{c}
   \nonumber \\
   & & \quad
       + \left( H + {\dot F \over 2 F} \right) {3 a \over 2 F}
       \mu_c v_{c}
       \Bigg]
       - {1 \over 2F} {a \over k} \mu_c v_c,
   \\
   & & {H + {\dot F \over 2 F} \over aF}
       \left( {aF \over H + {\dot F \over 2 F}} \Psi \right)^{\displaystyle\cdot}
       = {{3 \dot F^2 \over 2F} + f_{,X} X \over
       \dot F + 2 H F} \Phi
   \nonumber \\
   & & \quad
       + {\mu_c \over \dot F + 2 H F} \Psi
       - {1 \over 2F} \mu_c \left( {a \over k} v_c
       - \chi \right),
\eea where \bea
   & & \Phi \equiv \varphi_{\delta F}, \quad
       \Psi \equiv \varphi_\chi
       + {\delta F_\chi \over 2 F}.
\eea In the context of $f(R)$ gravity we simply set $X \equiv 0$. In
the absence of the CDM component, these equations are presented in
Eqs.\ (85) and (86) of \cite{HN-GGT-2005}. For the CDM part, from
Eq.\ (\ref{eq-6i}) and Eq.\ (\ref{eq-7i}), respectively, we have
\bea
   & & \dot \delta_{c}
       =
       - {k \over a} v_{c}
       + \kappa
       - 3 H \alpha,
   \\
   & & \dot v_{c}
       + H v_{c}
       = {k \over a} \alpha,
\eea where from Eqs.\ (\ref{eq-1})-(\ref{eq-3}) we have \bea
   & & \alpha
       = {1 \over H + {\dot F \over 2 F}}
       \left( \dot \Phi
       + {1 \over 2F} {a \over k} \mu_c v_{c} \right),
   \\
   & & \kappa
       = {1 \over H + {\dot F \over 2 F}}
       \Bigg[ {k^2 \over a^2} \Phi
       - {1 \over 2 F} \delta \mu_{c}
   \nonumber \\
   & & \quad
       - {1 \over 2 F} \left(
        3 H \dot F
        - f_{,X} X - 2 f_{,XX} X^2 \right) \alpha
        \Bigg],
   \\
   & & \chi
       = {a^2 \over k^2} \left(
       - {3 \over 2 F} {a \over k} \mu_c v_{c}
       + \kappa
       + {3 \over 2} {\dot F \over F} \alpha \right).
\eea These equations give a set of differential equations for
$\Phi$, $\Psi$, $\delta_c$ and $v_c$; these can be combined to give
a fourth-order differential equation. $\Phi$ and $\Psi$ are
gauge-invariant combinations, and all the other perturbation
variables are evaluated in the UFG; $\delta_c$ and $v_c$ evaluated
in the UFG are equivalent to gauge-invariant combinations $\delta_{c
\delta F} \equiv \delta_c + (3 H \dot F) \delta F$ and $v_{c \delta
F} \equiv v_c - (1/a \dot F) \delta F$, respectively. The CDM
density perturbation in the CDM-comoving gauge can be constructed by
\bea
   & & \delta_{c v_c} \equiv \delta_c + 3 H {a \over k} v_c,
\eea where the right-hand-side can be evaluated in the UFG.

\subsection*{Appendix C.3: CDM-comoving gauge}

The CDM-comoving gauge (CCG) takes $v_c \equiv 0$ as the temporal
gauge condition. From Eq.\ (\ref{eq-7i}) we have $\alpha = 0$.
Equations (\ref{eq-6i}) and (\ref{eq-3}), Eq.\ (\ref{eq-4}), and
Eqs.\ (\ref{eq-1}) and Eq.\ (\ref{eq-3}), respectively, give \bea
   & & \dot \delta_c
       = {k^2 \over a^2} \chi
       - {3 \over 2F} \left(
       {1 \over 2} f_{,X} \dot \phi \delta \phi
       - \delta \dot F
       + H \delta F \right),
   \label{eq-1-CCG} \\
   & & \dot \chi + \left( H + {\dot F \over F} \right) \chi
       = \varphi + {\delta F \over F},
   \label{eq-2-CCG} \\
   & & \dot \varphi
       = {1 \over F} \left(
       {1 \over 2} f_{,X} \dot \phi \delta \phi
       - \delta \dot F
       + H \delta F \right).
   \label{eq-3-CCG}
\eea In order to complete we need a first-order differential
equation for $\delta F$. From Eqs.\ (\ref{eq-2}) and (\ref{eq-3}) we
have \begin{widetext} \bea
   & & {1 \over 2} \left( f_{,X} + 2 f_{,XX} X \right)
       \dot \phi \delta \dot \phi
       - {3 \dot F \over 2 F} \delta \dot F
       + {1 \over 2} \left[ f_{,\phi}
       - 2 f_{,\phi X} X
       + 3 \left( H + {\dot F \over 2F} \right) f_{,X} \dot \phi
       \right] \delta \phi
   \nonumber \\
   & & \quad
       + \left[ 3 H^2 + {\mu_c \over F}
       - {3 \dot F \over 2F} H
       - {1 \over F} \left(
       {1 \over 2} f
       - f_{,X} X \right)
       + {k^2 \over a^2} \right] \delta F
       = - 2 F {k^2 \over a^2} \varphi
       + \delta \mu_c
       + \left( \dot F + 2 H F \right) {k^2 \over a^2} \chi,
   \label{eq-4-CCG}
\eea \end{widetext} where we have $\delta F = F_{,\phi} \delta \phi$
for (i), and $\delta \phi = 0 = X$ for (ii). Equations
(\ref{eq-1-CCG})-(\ref{eq-4-CCG}) provide a complete set of
equations for $\delta_c$, $\chi$, $\varphi$ and $\delta F$.

Equations (\ref{eq-1-CCG})-(\ref{eq-3-CCG}) can be combined to give
\bea
   & & {1 \over a^2} \left[ a^2 \left( \delta_c + 3 \varphi \right)^{\displaystyle\cdot}
       \right]^{\displaystyle\cdot}
       = - {k^2 \over a^2} \alpha_\chi,
   \label{ddot-delta-eq}
\eea where $\delta_c + 3 \varphi \equiv \delta_{c \varphi} \equiv 3
\varphi_{\delta_c}$ is a gauge-invariant combination, thus can be
evaluated in any gauge condition; $\delta_{c \varphi}$ is $\delta_c$
in the uniform-curvature gauge ($\varphi \equiv 0$) and
$\varphi_{\delta_c}$ is the spatial curvature perturbation $\varphi$
in the uniform-CDM-density gauge ($\delta_c \equiv 0$).

In the case of $f(R)$ gravity we suggest the following alternative
form. From $\delta R = \delta \mu - 3 \delta p$, Eq.\ (\ref{eq-5}),
and Eq.\ (\ref{eq-6i}) we have \begin{widetext} \bea
   & & \delta \ddot F
       + 3 H \delta \dot F
       + \left(
       - {1 \over 3} R
       + {k^2 \over a^2} \right)
       {\delta F}
       + {1 \over 3} F \delta R
       = \dot F \kappa
       + {1 \over 3} \left( \delta \mu_m - 3 \delta p_m \right),
   \label{fR-dot-F-eq} \\
   & & \dot \kappa
       + \left( 2 H - {\dot F \over F} \right) \kappa
       =
       - 3 H {\delta \dot F \over F}
       + \left( {1 \over 2} f
       + 3 H \dot F
       - \mu_m
       - F {k^2 \over a^2} \right) {\delta F \over F^2}
       - {1 \over 2} \delta R
       + {1 \over F} \delta \mu_m,
   \\
   & & \dot \delta_c = \kappa.
   \label{fR-dot-CDM-eq}
\eea For a CDM component we have $\mu_m = \mu_c$, $\delta \mu_m =
\mu_c \delta_c$, and $\delta p_m = 0$.
\end{widetext}

\vskip .5cm
%
%
\acknowledgments

We thank Dr.\ Yong-Seon Song for useful discussions. H.N.\ was
supported by Mid-career Research Program through National Research
Foundation funded by the MEST (No.\ 2010-0000302). J.H.\ was
supported by Korea Research Foundation Grant funded by the Korean
Government (KRF-2008-341-C00022).

%
%



\begin{thebibliography}{99}
\bibitem{DE-review-2010}
         S. Tsujikawa, arXiv:1004.1493v1; D. Sapone, arXiv:1006.5694.
\bibitem{Acquaviva-etal-2008}
         V. Acquaviva, A. Hajian, D.N. Spergel and S. Das,
         Phys.\ Rev.\  D {\bf 78}, 043514 (2008)
         [arXiv:0803.2236 [astro-ph]].
\bibitem{Amendola-etal-2008}
         L. Amendola, M. Kunz and D. Sapone,
         JCAP {\bf 0804}, 013 (2008)
         [arXiv:0704.2421 [astro-ph]].
\bibitem{Bean-2009}
         R. Bean, arXiv:0909.3853 [astro-ph.CO];
         R. Bean and M. Tangmatitham,
         Phys.\ Rev.\  D {\bf 81}, 083534 (2010)
         [arXiv:1002.4197 [astro-ph.CO]].
\bibitem{Bertschinger-Zukin-2008}
         E. Bertschinger and P. Zukin,
         Phys.\ Rev.\  D {\bf 78}, 024015 (2008)
         [arXiv:0801.2431 [astro-ph]].
\bibitem{Beynon-etal-2009}
         E. Beynon, D.J. Bacon and K. Koyama,
         Mon.\ Not.\ Roy.\ Astron.\ Soc.\  {\bf 403}, 353 (2010)
         [arXiv:0910.1480 [astro-ph.CO]].
\bibitem{Caldwell-etal-2007}
         R. Caldwell, A. Cooray and A. Melchiorri,
         Phys.\ Rev.\  D {\bf 76}, 023507 (2007)
         [arXiv:astro-ph/0703375].
\bibitem{Cui-etal-2010}
         W. Cui, P. Zhang and X. Yang,
         Phys.\ Rev.\  D {\bf 81}, 103528 (2010)
         [arXiv:1001.5184 [astro-ph.CO]].
\bibitem{Daniel_etal-2010}
         S.F. Daniel, E.V. Linder, T.L. Smith, R.R. Caldwell, A. Cooray,
         A. Leauthaud, and L. Lombriser, \prd \textbf{81}, 123508 (2010)
         [arXiv:1002.1962 [astro-ph.CO]].
\bibitem{Daniel-etal-2008}
         S.F. Daniel, R.R. Caldwell, A. Cooray and A. Melchiorri,
         Phys.\ Rev.\  D {\bf 77}, 103513 (2008)
         [arXiv:0802.1068 [astro-ph]].
\bibitem{Daniel-etal-2009}
         S.F. Daniel {\it et al.},
         Phys.\ Rev.\  D {\bf 80}, 023532 (2009)
         [arXiv:0901.0919 [astro-ph.CO]].
\bibitem{DeFelice-Suyama-2010}
         A. De Felice and T. Suyama,
         arXiv:1010.3886 [astro-ph.CO].
\bibitem{DeFelice-etal-2010}
         A. De Felice, S. Mukohyama and S. Tsujikawa,
         Phys.\ Rev.\  D {\bf 82}, 023524 (2010)
         [arXiv:1006.0281 [astro-ph.CO]].
\bibitem{Dore-etal-2007}
         O. Dor\'e {\it et al.},
         arXiv:0712.1599 [astro-ph].
\bibitem{Ferreira-2010}
         P.G. Ferreira and C. Skordis,
         Phys.\ Rev.\  D {\bf 81}, 104020 (2010)
         [arXiv:1003.4231 [astro-ph.CO]].
\bibitem{Giannantonio-etal-2009}
         T. Giannantonio, M. Martinelli, A. Silvestri and A. Melchiorri,
         JCAP {\bf 1004}, 030 (2010)
         [arXiv:0909.2045 [astro-ph.CO]].
\bibitem{Guzik-etal-2009}
         J. Guzik, B. Jain and M. Takada,
         Phys.\ Rev.\  D {\bf 81}, 023503 (2010)
         [arXiv:0906.2221 [astro-ph.CO]].
\bibitem{Hu-2008}
         W. Hu,
         Phys.\ Rev.\  D {\bf 77}, 103524 (2008)
         [arXiv:0801.2433 [astro-ph]].
\bibitem{Hu-Sawicki-2007}
         W. Hu and I. Sawicki,
         Phys.\ Rev.\  D {\bf 76}, 104043 (2007)
         [arXiv:0708.1190 [astro-ph]].
\bibitem{Huterer-Linder-2007}
         D. Huterer and E.V. Linder,
         Phys.\ Rev.\  D {\bf 75}, 023519 (2007)
         [arXiv:astro-ph/0608681].
\bibitem{Jain-Zhang-2008}
         B. Jain and P. Zhang,
         Phys.\ Rev.\  D {\bf 78}, 063503 (2008)
         [arXiv:0709.2375 [astro-ph]].
\bibitem{Laszlo-Bean-2008}
         I. Laszlo and R. Bean,
         Phys.\ Rev.\  D {\bf 77}, 024048 (2008)
         [arXiv:0709.0307 [astro-ph]].
\bibitem{Linder-Cahn-2007}
         E.V. Linder and R.N. Cahn,
         Astropart.\ Phys.\  {\bf 28}, 481 (2007)
         [arXiv:astro-ph/0701317].
\bibitem{Martinelli-etal-2010}
         M. Martinelli, E. Calabrese, F. De Bernardis, A. Melchiorri,
         L. Pagano and R. Scaramella,
         arXiv:1010.5755 [astro-ph.CO].
\bibitem{Nesseris-2007}
         S. Nesseris and L. Perivolaropoulos,
         Phys.\ Rev.\  D {\bf 77}, 023504 (2008)
         [arXiv:0710.1092 [astro-ph]].
\bibitem{Pogosian_etal-2010}
         L. Pogosian, A. Silvestri, K. Koyama and G.B. Zhao,
         Phys.\ Rev.\  D {\bf 81}, 104023 (2010)
         [arXiv:1002.2382 [astro-ph.CO]].
\bibitem{Schmidt-2008}
         F. Schmidt,
         Phys.\ Rev.\  D {\bf 78}, 043002 (2008)
         [arXiv:0805.4812 [astro-ph]].
\bibitem{Serra-etal-2009}
         P. Serra, A. Cooray, S.F. Daniel, R. Caldwell, and A. Melchiorri,
         Phys.\ Rev.\ D {\bf 79}, 101301 (2009)
         [arXiv:0901.0917 [astro-ph.CO]].
\bibitem{Shapiro-etal-2010}
         C. Shapiro, S. Dodelson, B. Hoyle, L. Samushia, B. Flaugher,
         arXiv:1004.4810v2 [astro-ph.CO].
\bibitem{Skordis-2009}
         C. Skordis,
         Phys.\ Rev.\ D {\bf 79}, 123527 (2009)
         [arXiv:0806.1238 [gr-gc]]
\bibitem{Song-Koyama-2008}
         Y.S. Song and K. Koyama,
         JCAP {\bf 0901}, 048 (2009)
         [arXiv:0802.3897 [astro-ph]].
\bibitem{Song-Dore-2009}
         Y.S. Song and O. Dore,
         JCAP {\bf 0903}, 025 (2009)
         [arXiv:0812.0002 [astro-ph]].
\bibitem{Song-etal-2010-01}
         Y.S. Song, L. Hollenstein, G. Caldera-Cabral and K. Koyama,
         JCAP {\bf 1004}, 018 (2010)
         [arXiv:1001.0969 [astro-ph.CO]].
\bibitem{Song_etal-2010-11}
         Y.S. Song, G.B. Zhao, D. Bacon, K. Koyama, R.C. Nichol and L. Pogosian,
         arXiv:1011.2106 [astro-ph.CO].
\bibitem{Tsujikawa-2007}
         S. Tsujikawa,
         Phys.\ Rev.\  D {\bf 76}, 023514 (2007)
         [arXiv:0705.1032 [astro-ph]].
\bibitem{Tsujikawa-etal-2008}
         S. Tsujikawa, K. Uddin, S. Mizuno, R. Tavakol, and J. Yokoyama,
         Phys.\ Rev.\ D \textbf{77}, 103009 (2008)
         [arXiv:0803.1106 [astro-ph]].
\bibitem{Uzan-2010}
         J.P. Uzan,
         Gen.\ Rel.\ Grav.\  {\bf 42}, 2219 (2010)
         [arXiv:0908.2243 [astro-ph.CO]].
\bibitem{Zhang-etal-2008}
         P. Zhang, R. Bean, M. Liguori and S. Dodelson,
         arXiv:0809.2836 [astro-ph].
\bibitem{Zhang-etal-2007}
         P. Zhang, M. Liguori, R. Bean and S. Dodelson,
         Phys.\ Rev.\ Lett.\  {\bf 99} (2007) 141302
         [arXiv:0704.1932 [astro-ph]].
\bibitem{Zhao-etal-2010}
         G.B. Zhao {\it et al.},
         Phys.\ Rev.\  D {\bf 81}, 103510 (2010)
         [arXiv:1003.0001 [astro-ph.CO]].
\bibitem{Zhao-etal-2009}
         G.-B. Zhao, L. Pogosian, A. Silvestri, and J. Zylberberg,
         Phys.\ Rev.\ D {\bf 79}, 083513 (2009)
         [arXiv:0809.3791v2 [astro-ph]];
         G.-B. Zhao, L. Pogosian, A. Silvestri and J. Zylberberg,
         Phys.\ Rev.\ Lett.\  {\bf 103} (2009) 241301
         [arXiv:0905.1326 [astro-ph.CO]].
\bibitem{Bardeen-1980}
         J.M. Bardeen, Phys. Rev. D \textbf{22}, 1882 (1980).
\bibitem{PN}
         J. Hwang and H. Noh, D. Puetzfeld, J. Cosmol. Astropart. Phys.
                {\bf 03}, 010 (2008).
\bibitem{fR-2010}
         C. Park, J. Hwang, H. Noh, Phys. Rev. D submitted (2010).
\bibitem{Amanullah-etal-2010}
         R. Amanullah, et al., Astrophy. J. \textbf{716}, 712 (2010).
\bibitem{Samushia-Ratra-2006}
         L. Samushia and B. Ratra, Astrophy. J. \textbf{650}, L5 (2006).
\bibitem{Vikhlinin-etal-2009}
         A. Vikhlinin, et al., arXiv:0903.5320.
\bibitem{Park-etal-PRL-2009}
         C. Park, J. Hwang, J. Lee, H. Noh, Phys. Rev. Lett. \textbf{103},
         151303 (2009).
\bibitem{Bardeen-1988}
         J.M. Bardeen, {\it Particle Physics and Cosmology}, edited by
                       L. Fang and A. Zee (Gordon and Breach, London, 1988), p1.
\bibitem{HN-GGT-2005}
         J. Hwang and H. Noh, Phys. Rev. D \textbf{71}, 063536 (2005).
\bibitem{HN-GGT-1996}
         J. Hwang and H. Noh, Phys. Rev. D \textbf{54}, 1460 (1996).
\end{thebibliography}
\end{document}